\documentclass{webofc}
\usepackage[varg]{txfonts}   
\usepackage[english]{babel}
\usepackage{here}
\usepackage{subfloat}
\usepackage{fontenc}
\usepackage[utf8]{inputenc}
\usepackage[colorlinks=false]{hyperref}
\usepackage[printonlyused]{acronym}
\usepackage{bm}
\usepackage{bbold}
\usepackage{amsmath}
\usepackage{slashed}
\usepackage{amssymb}
\usepackage{textcomp}
\usepackage[usenames]{color}
\usepackage{enumerate}
\usepackage{dcolumn}
\usepackage{longtable}
\usepackage[usenames]{color}
\newcommand{\be}{\begin{equation}}
\newcommand{\ee}{\end{equation}}
\newcommand{\bea}{\begin{eqnarray}}
\newcommand{\eea}{\end{eqnarray}}
\newcommand{\ba}{\begin{array}}
\newcommand{\ea}{\end{array}}
\newcommand{\bi}{\begin{itemize}}
\newcommand{\ei}{\end{itemize}}

\begin{document}
\title{SU(3)$_f$ Constraints on Hypernuclear Energy Density Functionals}
%
%

\author{\firstname{Horst} \lastname{Lenske}\inst{1}\fnsep\thanks{\email{horst.lenske@physik.uni-giessen.de}} \and
        \firstname{Madhumita} \lastname{Dhar}\inst{2}\fnsep\thanks{\email{madhumita.dhar@cgec.org.in}}
}

\institute{Institut f\"{u}r Theoretische Physik, Justus-Liebig-Universit\"at Gie\ss en, D-35392 Gie\ss en, Germany
\and
           Cooch Behar Government Engineering College, Cooch Behar, West Bengal, India 736170
          }

\abstract{
A covariant hypernuclear energy density functional (EDF) is derived from in--medium nucleon--meson vertex functionals, assuring the proper description of nuclear mean--field dynamics. The fundamental SU(3) coupling constants for the mean--field relevant vector ($m=V$) and scalar ($m=S$) interactions as functionals of the total baryon density $\rho_B$ are determined. Scalar and vector potentials and the resulting hyperon
mean--fields in asymmetric nuclear matter are constructed and discussed, addressing also effects from 3--body interactions. $\Lambda$--$\Sigma^0$ mixing in asymmetric nuclear through the coupling to the background isovector mean--field is addressed.
}
\maketitle
\section{Introduction}\label{sec:intro}

With the focus on mean--field dynamics and the description of bulk properties like single particle separation and total binding energies, energy density functionals (EDF) have been developed for hypernuclear and neutron star studies  \cite{Glendenning:1991es,GL:1992,RMFHyp1,SkHyp1,Kanakis-Pegios:2020kzp}, the latter often with special focus on the still unsolved so--called hyperon puzzle, limiting the maximal masses of neutron stars.  Covariant Lagrangian approaches and non--relativistic Skyrme-type EDFs were constructed in parallel. In order to overcome the scarce data base, Skyrme EDFs were derived from G--matrix interactions \cite{Lanskoy:1997xq,Schulze:2014oia}, still being used  and recently applied in investigations of light Cascade--nuclei \cite{Guo:2021vsx}. First attempts to implement SU(3) symmetry into hypernuclear physics by using $\chi$EFT  were made already more than a decade ago, see e.g. \cite{Finelli:2009tu}.

The standard approach to microscopic SU(3)--based in--medium dynamics follows closely the rather successful strategies developed over the last decades for pure nucleon dynamics, namely first fixing free space two--body interactions which in a second step are inserted into an in--medium scheme like Brueckner G--matrix theory, eventually extended by 3--body forces, e.g. \cite{Soma:2008nn,Holt:2019bah}. The resulting in--medium scattering amplitudes are used to construct mean--field self--energies given by static potentials, depending on momentum if non--localities due to anti--symmetrization are treated explicitly.  That is also the strategy adopted in Density Dependent Relativistic Hadron (DDRH) theory, developed some time ago at Giessen (see \cite{Lenske:2004,Lenske:2018bgr,Lenske:2018bvq} and further references therein)
and used here as an entry point for SU(3) mean--field dynamics.

\begin{figure}[h]
\centering
\includegraphics[width=13cm,clip]{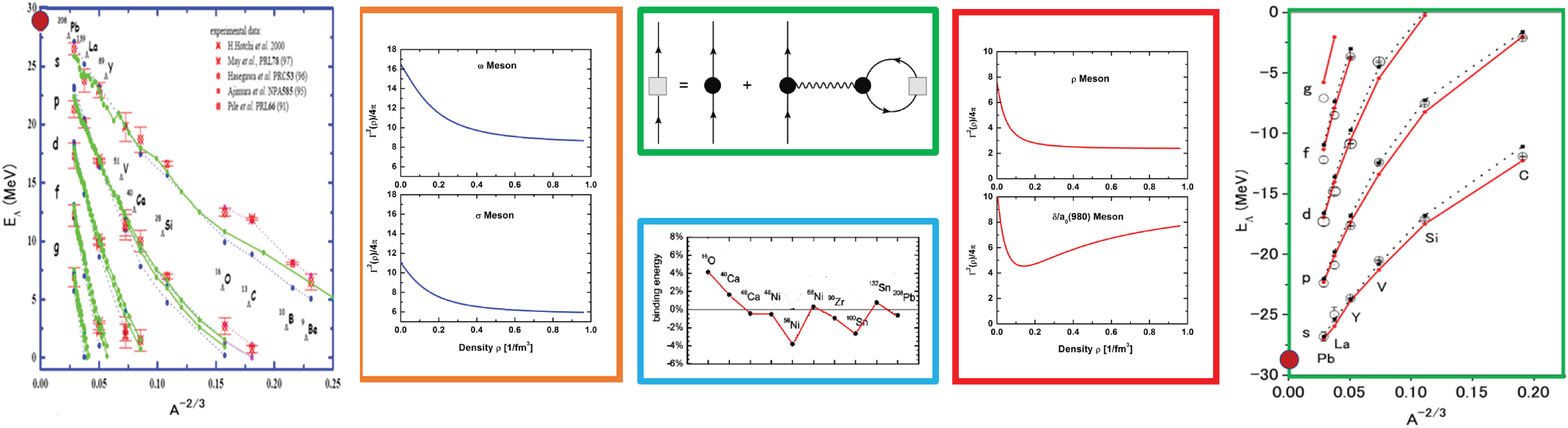}
\caption{DDRH vertices derived from DBHF self--energies. The diagram in the upper center represents the dressed vertices (boxes) as given by the bare NN--meson couplings (filled circles). Baryon and meson propagators are indicated by straight and wavy lines, respectively.
The resulting isoscalar and isovector coupling functionals are shown at the left ($\omega,\sigma$) and the right ($\rho,\delta/a_0(980)$) , respectively. Applications to stable nuclei over the mass chart are illustrated in the lower center by displaying the relative deviations of calculated and experimental binding energies. Also shown are mean--field results for single--$\Lambda$ separation energies: on the far left covariant DDRH results and on the very right non--relativistic results obtained by a folding approach using the G--matrix from the ESC-08 interaction \cite{Yamamoto:2014jga} are shown. For further details see Ref. \cite{Lenske:2018bgr}. }
\label{fig:DDRH}
\end{figure}

The DDRH scheme is based on covariant Dirac--Brueckner--Hartree--Fock (DBHF) G--matrix calculations. The DBHF self--energies are projected to an effective Hartree theory. For practical calculations, effective density--dependent NN--meson coupling functionals are derived such that the DBHF self--energies are reproduced. The functionals, incorporating naturally anti--symmetrization contributions, serve to recast the DBHF results into a covariant EDF. The interaction part is given by the mean--field producing isoscalar ($\sigma,\omega$) and isovector ($\delta/a_0(980),\rho$) meson fields. Applying the variational rules of covariant density functional theory (DFT), Dirac equations for baryons and (classical) Klein--Gordan equations for the static meson fields have to be solved. The approach and results are summarized in Fig. \ref{fig:DDRH} where the vertex equation is illustrated graphically and the four NN--meson functionals are shown together with results for nuclear binding energies of stable nuclei all over the mass table. Results for infinite nuclear matter are found in Tab.\ref{tab:EoS}. The DDRH vertices were used, in fact, also in investigations of single--$\Lambda$ nuclei \cite{Keil:1999hk,Keil:2002ad} by using a simple scaling approach, widely used at that time. A look into the original research paper \cite{Keil:1999hk} will be quite instructive for understanding of the uncertainties of hypernuclear parameters at that time, which in fact are still waiting for experimental support. Also in Fig. \ref{fig:DDRH} these early results for $\Lambda$--separation energies are compared to data and corresponding results by a G--matrix folding approach using the non--relativistic Nijmegen--interaction \cite{Yamamoto:2014jga}.

In the following, we take a fresh view on the in--medium physics of octet baryons. Although the use of SU(3) relations is a standard tool in octet physics and beyond, the new aspect of the approach presented below is to exploit those relations not on the Born--level but on the level of the already fully resummed diagrams of nucleon in--medium interactions. The four NN--vertex functionals, available from DDRH theory, are sufficient to derive the mean--field producing in--medium interactions for the full set of octet baryons.

\section{In--Medium SU(3) Density Functional Theory}\label{sec:sec-1}
\subsection{Formal Aspects of SU(3) Mean--Field Theory}
The purpose of this section is to give an overview of the construction of a Lagrangian density describing octet baryons and their interactions with the meson nonets. Interactions are defined by the SU(3) rules introduced by de Swart \cite{deSwart:1963pdg}. Denoting the eight Gell--Mann SU(3) matrices by $\lambda^a$, the baryons are represented as usual by the 3--by--3 matrix $\mathcal{B}=\frac{1}{\sqrt{2}}\sum^8_{a=1}\lambda^a B^a$, see e.g. \cite{Lenske:2018bgr}. The octet meson matrices are defined accordingly, e.g. $\mathcal{V}_8=\frac{1}{\sqrt{2}}\sum^8_{a=1}\lambda^a V^a$ for the vector mesons with $J^\pi=1^-$ . Singlet mesons are included by the 3--by--3 unity matrix $\lambda^0$, e.g. $\mathcal{V}_1=\lambda^0V^0$ where in the vector case $V^0=\phi_1$ is given by the (unmixed) singlet vector meson $\phi_1$. Scalar ($J^\pi=0^+$) and pseudo--scalar ($J^\pi=0^-$) mesons are described accordingly.

We obtain the Lagrangian density
$\mathcal{L}=\mathcal{L}_{B}+\mathcal{L}_{M}+\mathcal{L}_{BBM}$.
The baryon and meson Lagrangians $\mathcal{L}_{B,M}$ are superpositions of standard Dirac-- and Klein--Gordan--type Lagrangians for the free motion of fermions and bosons, respectively. SU(3) symmetry breaking is included by using physical hadron masses. For finite systems, electromagnetic interactions are supplemented.

Of central interest for the theory is the baryon--meson interaction Lagrangian given by the sum of Lagrangians describing the three classes of interactions, $\mathcal{L}_{BBM}=\sum_{m=P,S,V}\mathcal{L}^{(m)}_{int}$.
The interaction Lagrangians are of the same formal structure, e.g. for the vector case.:
\be
\mathcal{L}^{(V)}_{int}=-\sqrt{2}\left(g^{(V)}_D \left[\overline{\mathcal{B}}\mathcal{BV}_8\right]_D+
g^{(V)}_F \left[\overline{\mathcal{B}}\mathcal{BV}_8\right]_F \right)-g^{(V)}_S\frac{1}{\sqrt{3}}\left[\overline{\mathcal{B}}\mathcal{BV}_1\right]_S
\ee
\be
\left[\overline{\mathcal{B}}\mathcal{BV}_8\right]_D=tr\left(\left\{\overline{\mathcal{B}},\mathcal{B} \right\}\mathcal{V}_8 \right) ; \quad
\left[\overline{\mathcal{B}}\mathcal{BV}_8\right]_F=tr\left(\left[\overline{\mathcal{B}},\mathcal{B} \right]\mathcal{V}_8 \right)
 ; \quad
\left[\overline{\mathcal{B}}\mathcal{BV}_1\right]_S=tr\left(\overline{\mathcal{B}}\mathcal{B} \right)tr\left(\mathcal{V}_1 \right).
\ee
where the $D$ and $F$ couplings are given by the baryon anti--commutator and the baryon commutator, respectively:

\subsection{SU(3) Mean--Field Theory}
The guiding principles are that firstly by the fit to scattering data NN--interactions fully inherit SU(3) symmetry, but e.g. by the use of physical masses incorporate also symmetry breaking effects, and secondly
that both free space and in-medium Bethe-Salpeter equations conserve the fundamental symmetries.
Thirdly, we note in addition that this is also true for the vertex equations by which the coupling functionals are derived from the DBHF G--matrix interactions. Thus, we formulate the  \emph{SU(3) DFT--Program}:
For each of the nonet interaction channels, pseudo--scalar (P), scalar (S), and vector (V) meson exchange, only three physical couplings are sufficient to fix the set of fundamental SU(3) couplings $\{g^{(m)}_D,g^{(m)}_F,g^{(m)}_S\}_{|m=P,S,V}$. If those are known, the full set of vertex functionals for the remaining baryon--meson couplings are determined. In the forthcoming, only the mean--field relevant scalar and vector parts will be discussed.

\begin{table}
  \centering
\begin{tabular}{|c|c|c|c|c|c|c|}
        \hline
     Model  &  $\rho_{sat}$/$fm^{-3}$   &    $E/A$/MeV  &   $K_\infty$/MeV  &    $E_{sym}$/MeV  &    $L_{sym}$/MeV  &    $K_{sym}$/MeV\\
     \hline
    2-body  &  0.180 & -15.603 & 281.945 &  31.154 &  88.627 & 201.399\\
    3-body  &  0.160 & -16.000 & 283.136 &  32.000 &  90.000 & 133.502\\
        \hline
      \end{tabular}
  \caption{Properties of the DDRH equation of state of infinite symmetric nuclear matter, obtained with 2--body interactions only (first row) and by adding 3--body interactions (second row). }\label{tab:EoS}
\end{table}

The full treatment of SU(3) baryon octet/meson nonet dynamics requires to consider octet--singlet meson mixing. For simplicity, but not as limitation by principle, we assume throughout ideal mixing, i.e. the charged--neutral octet mesons ($\eta,\sigma,\omega$) do not contain $s\bar s$--components while their physical singlet partners ($\eta'(960),\sigma'\sim f_0(980),\phi(1020)$) are pure $s\bar s$ configurations. These constraints are fulfilled by the ideal mixing angle $\tan{\theta}=1/\sqrt{2}$ ($\theta\sim 35.26^\circ$) which is common for all three interaction sectors. For example, denoting the vector meson mixing angle by $\theta_V$ the vector SU(3) couplings are explicitly:
\bea\label{eq:gDFS}
g^{(V)}_D&=&\frac{1}{2}\left[3g_{NN\rho}-\sqrt{3}\left(g_{NN\omega}\sin{\theta_V}-
g_{NN\phi}\cos{\theta_V}  \right)  \right]\nonumber\\
g^{(V)}_F&=&\frac{1}{2}\left[g_{NN\rho}+\sqrt{3}\left(g_{NN\omega}\sin{\theta_V}-
g_{NN\phi}\cos{\theta_V}  \right)  \right]\\
g^{(V)}_S&=&\sqrt{2}\left[g_{NN\omega}\cos{\theta_V}+
g_{NN\phi}\sin{\theta_V}   \right],\nonumber
\eea
which -- together with the corresponding relation for the scalar (and pseudo--scalar) couplings -- are the key relations for the whole approach.
Compatible with ideal mixing is the hypothesis that nucleons do not couple to the (physical) singlet mesons. Taking that view, we fix $g_{NN\eta'}(\rho_B)=g_{NNf_0}(\rho_B)=g_{NN\phi}(\rho_B)\equiv 0$. By means of Eq.\eqref{eq:gDFS} and the two sets of scalar and vector DDRH Hartree functionals, the prerequisites are at hand for the derivation of the full set of scalar and vector SU(3) in--medium couplings $g^{(S,V)}_{D,F,S}(\rho_B)$. \footnote{Lorentz--invariance of the theory demands $\rho_B\equiv \sqrt{j_{B\mu}j_B^\mu}$ with the total baryon 4--current $j_{B\mu}=\sum_{b=N,Y}\overline{\Psi}_b\gamma_\mu\Psi_b$ which in the nuclear rest frame simplifies to the standard expression for the total baryon number density.} Thus, a SU(3)--based EDF can be constructed. However, numerically the values of the deduced SU(3) density--dependent coupling functionals will depend on the in--medium NN--interaction model.

\begin{figure}[h]
\centering
\includegraphics[width=10cm,clip]{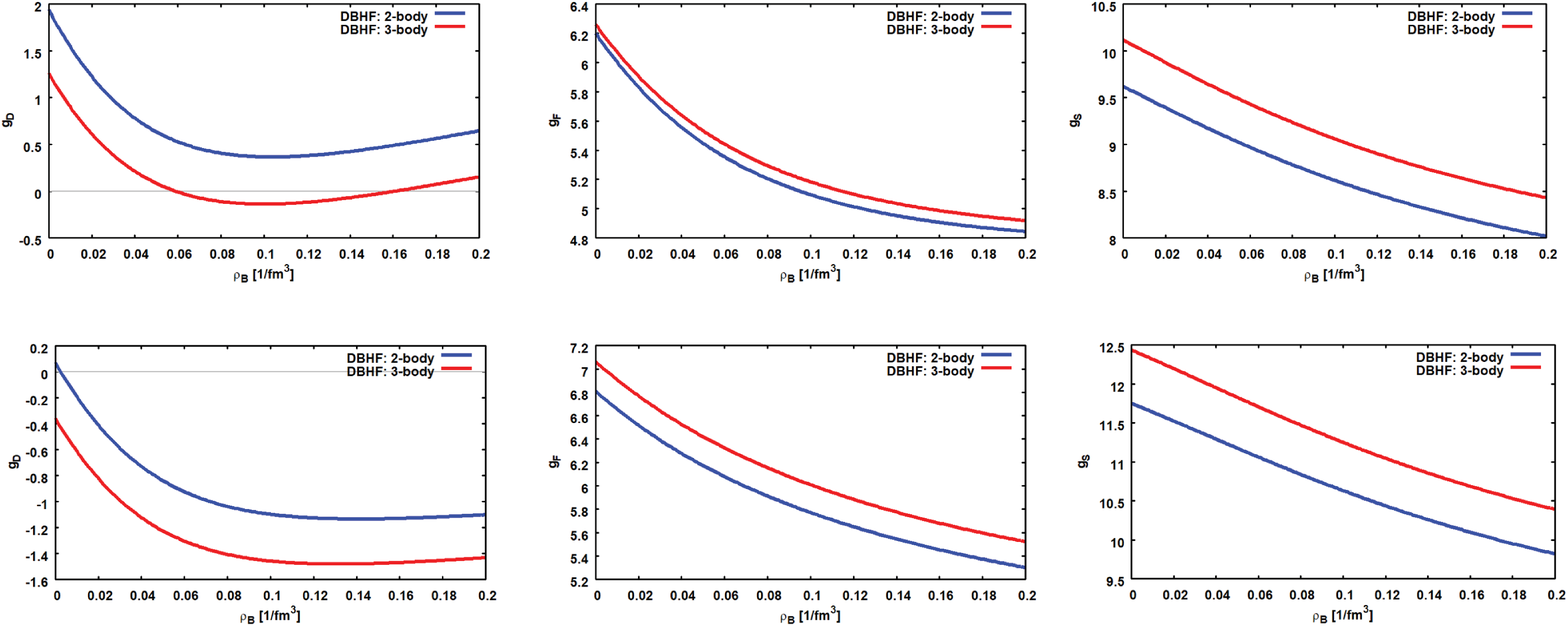}
\caption{Elementary SU(3) in--medium vertices $g_D$ (left), $g_F$ (center), $g_S$ (right)  for scalar (upper row) and vector interactions (lower row). As indicated, results are shown without (blue) and with three--body (red) interactions, the latter are of Fujita--Miyazawa type \cite{Pieper:2008rui} and act only among nucleons. }
\label{fig:gDFS}
\end{figure}

In Fig. \ref{fig:gDFS} the fundamental SU(3) coupling functionals for scalar and vector interactions are shown as a function of the baryon number density. The chosen scheme leads to a suppression of the D-couplings $g^{(S,V)}_D(\rho_B)$ such that mean--field dynamics are dominated by the octet F-couplings $g^{(S,V)}_F(\rho_B)$ and the singlet couplings $g^{(S,V)}_S(\rho_B)$. Similar results are found for other choices for the mixing angle, either the empirical linear or the Gell--Mann--Okuba mass relation (see also Ref. \cite{Adamuscin:2018zji}).

\begin{figure}[h]
\centering
\sidecaption
\includegraphics[width=8cm,clip]{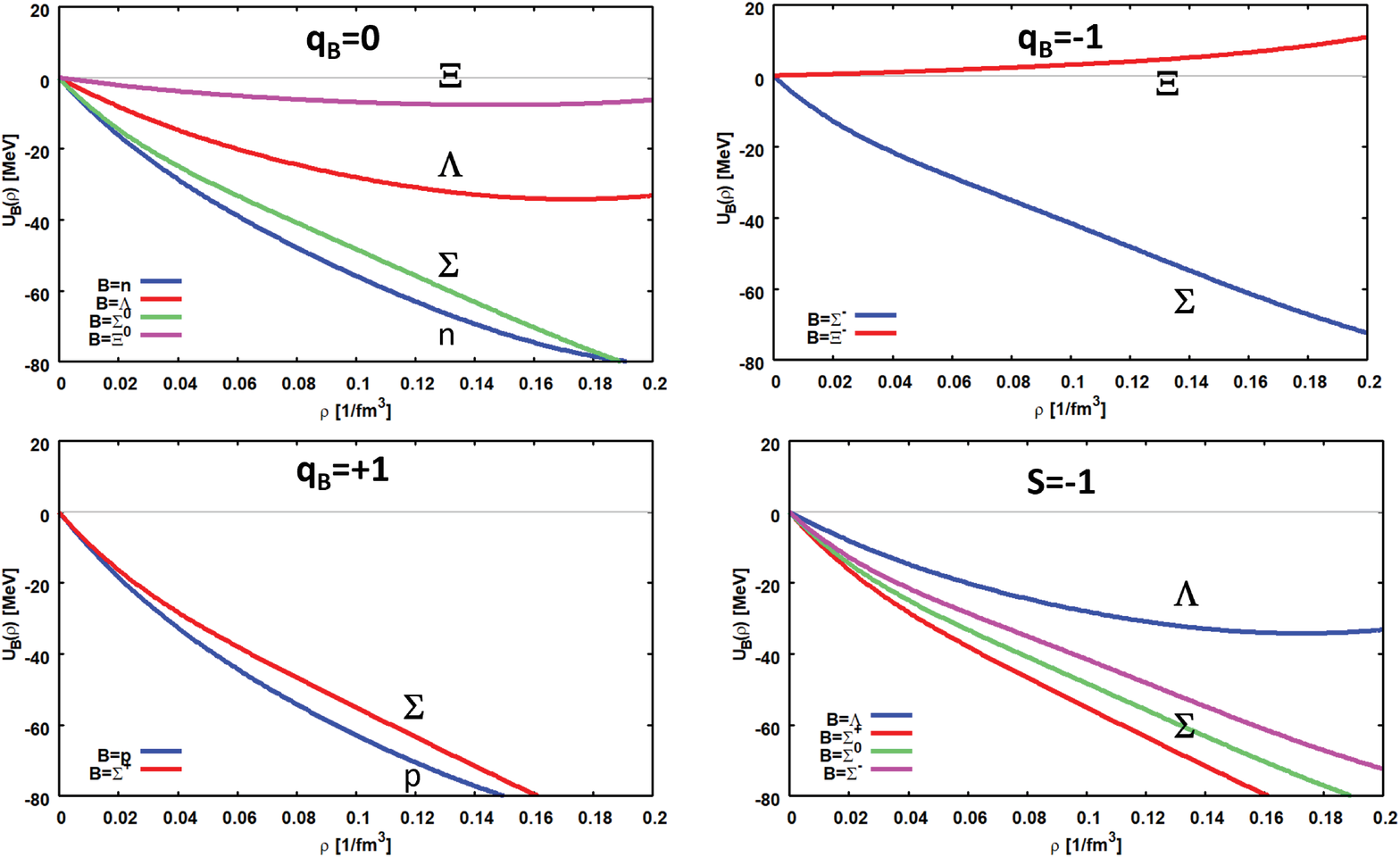}
\caption{Baryon covariant mean--fields in infinite asymmetric matter are shown where  $U_B(\rho)=U^{(S)}_B(\rho)+U^{(V)}_B(\rho)$ is the sum of scalar and vector potentials, including both isoscalar and isovector interactions. Results for $Z/A=0.4$ are displayed, realized e.g. in $^{10}$Be and approximately in $^{48}$Ca, $^{124}$Sn, and $^{208}$Pb. Results are shown for baryons of equal charge (left and right top) and compared for the $S=-1$ multiplet of hypercharge $Y=0$ consisting of the $\Sigma$ iso--triplet and the $\Lambda$ iso--singlet.  }
\label{fig:MeanF}
\end{figure}

\subsection{Mean--Fields of Octet Baryons}
The standard variational Euler--Lagrange rules serve to obtain the equations of motion, finally to be solved numerically.
As seen in Fig. \ref{fig:DDRH} and Tab. \ref{tab:EoS} the covariant DDRH--EDF leads already in calculations with only 2--body interactions to rather satisfactory results both for finite nuclei and infinite nuclear matter, much better than found in corresponding non--relativistic BHF studies.  The reason is that the underlying DBHF calculations include important field--theoretical aspects like the coupling to the negative--energy Fermi sea, acting like a 3--body interaction (the so--called polarization-- or Z--diagrams), causing e.g. the difference between Dirac scalar and vector densities. What is  missing, however, are the Fujita--Miyazawa 3--body contributions involving excitation of nucleon resonances, studied in the past in much detail e.g. by the Urbana group \cite{Pieper:2008rui}. Here, these were simulated by effective density dependent 2--body interactions \cite{Soma:2008nn} where the strengths were adjusted to reproduce the saturation properties of the  AV18+UIX--results of Akmal et al. \cite{NS3}. Overall modifications of scalar and vector NN--vertices on a level of about $5$\% or less are obtained which as seen in Tab. \ref{tab:EoS} are sufficient to adjust the equation of state closer to the empirical realm. However, as seen in Fig. \ref{fig:gDFS}, the nucleonic 3--body corrections have a pronounced influence on the SU(3) couplings, lowering $g^{(m)}_D$ considerably, compensated in part by increasing $g^{(m)}_{F,S}$.

Rather than considering separately the Dirac scalar and vector fields  $U^{(S,V)}_B$, more reasonable quantities are their sums $U_B= U^{(S)}_B+ U^{(V)}_B$ which are the leading order non--relativistic reductions. With additional correction terms \cite{Adamian:2021gnm}, they become finally Schroedinger--equivalent potentials. The potentials $U_B$ are displayed in Fig. \ref{fig:MeanF} for infinite asymmetric nuclear matter with charge--mass ratio $Z/A=0.4$, as encountered in $^{10}$Be approximately in nuclei up to $^{208}$Pb.
Remarkably, the $\Lambda$--potential agrees almost perfectly well with the potential derived much earlier in Ref. \cite{Keil:1999hk} by a $\chi^2$--fit to single-$\Lambda$ separation energies, see Fig. \ref{fig:DDRH}.

\begin{figure}
\centering
\includegraphics[width=8cm,clip]{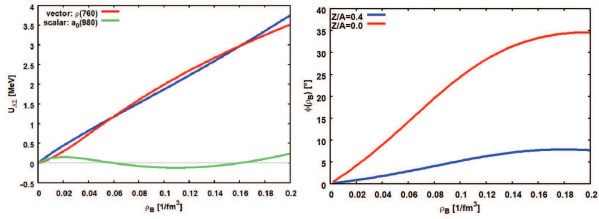}
\caption{Density dependence of the $\Lambda$--$\Sigma^0$ in--medium mixing potential (left) and the resulting mixing angles (right) in nuclear matter with $Z/A=0.4$ and in pure neutron matter, $Z/A=0$. The scalar and vector components of the mixing potential are indicated. Note the strong increase of mixing with decreasing proton content.}
\label{fig:LSmix}
\end{figure}

In closing the paper we point to the interesting fact that the approach naturally predicts $\Lambda$--$\Sigma^0$ mixing in asymmetric matter by the isovector mean--field. The mixing potential, depending on the charge asymmetry $Z/A$, and may be written as
\be
U_{\Lambda\Sigma}(\rho_B)=(1-2\frac{Z}{A})\left(U^{(S)}_{NN}(\rho_B)\frac{g_{\Lambda\Sigma \delta}}{g_{NN \delta}}
+U^{(V)}_{NN}(\rho_B)\frac{g_{\Lambda\Sigma \rho}}{g_{NN \rho}}\right)
\ee
where the scalar and vector isovector background potentials are denoted by $U^{(S,V))}_{NN}$ and $g_{\Lambda\Sigma m}=\sqrt{\frac{2}{3}}g^{(S,V)}_D$. In Fig. \ref{fig:LSmix} the potential is shown together with the mixing angles obtained in asymmetric matter with $Z/A=0.4$  and pure neuron matter, $Z/A=0$. We emphasize that this kind of mixing is a genuine many--body effect, different from the well known mixing on the level of 2--body matrix elements.

\section{Summary and Outlook}

Well constraint hypernuclear energy density functionals are highly demanded for exploratory investigations of medium-- and heavy--mass hypernuclear, allowing also safe extrapolations to neutron stars. For that goal, it was pointed out that well studied nuclear EDFs together with the general rules of SU(3) physics provide in principle a promising and appropriate entry point. Here, we have presented the formal aspects and derived vertex functionals relevant for hypernuclear mean--field dynamics. Applications to physical systems from hypernuclei of medium and heavy mass to neutrons stars are in preparation.

%
%

\end{document}